\documentclass[12pt]{article}
\usepackage{html}
\usepackage{epsf}
\usepackage[pdftex]{graphicx}
\usepackage{graphicx}
\usepackage{amssymb}
\usepackage{hyperref}
\begin{document}
\begin{titlepage}
\includegraphics[width=150mm]{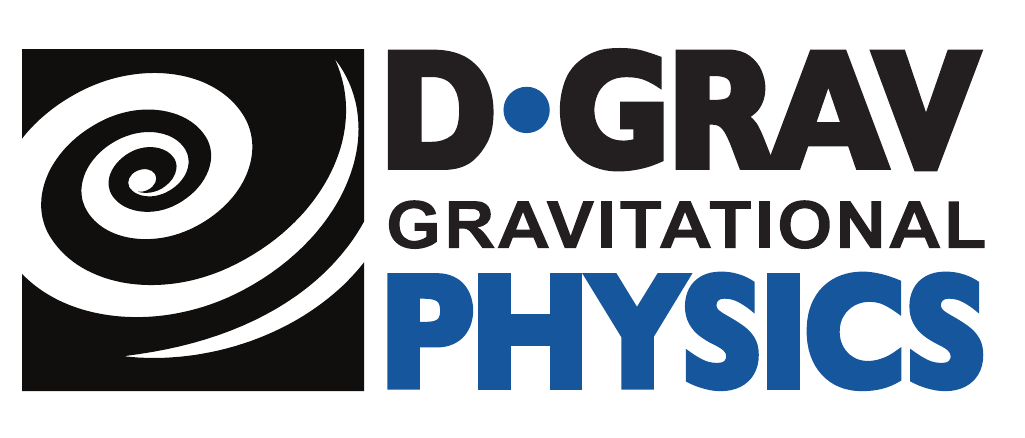}
\begin{center}
{ \Large {\bf MATTERS OF GRAVITY}}\\ 
\bigskip
\hrule
\medskip
{The newsletter of the Division of Gravitational Physics of the American Physical 
Society}\\
\medskip
{\bf Number 49 \hfill June 2017}
\end{center}
\begin{flushleft}
\tableofcontents
\end{flushleft}
\end{titlepage}
\vfill\eject
\begin{flushleft}
\section*{\noindent  Editor\hfill}
David Garfinkle\\
\smallskip
Department of Physics
Oakland University
Rochester, MI 48309\\
Phone: (248) 370-3411\\
Internet: 
\htmladdnormallink{\protect {\tt{garfinkl-at-oakland.edu}}}
{mailto:garfinkl@oakland.edu}\\
WWW: \htmladdnormallink
{\protect {\tt{http://www.oakland.edu/?id=10223\&sid=249\#garfinkle}}}
{http://www.oakland.edu/?id=10223&sid=249\#garfinkle}\\

\section*{\noindent  Associate Editor\hfill}
Greg Comer\\
\smallskip
Department of Physics and Center for Fluids at All Scales,\\
St. Louis University,
St. Louis, MO 63103\\
Phone: (314) 977-8432\\
Internet:
\htmladdnormallink{\protect {\tt{comergl-at-slu.edu}}}
{mailto:comergl@slu.edu}\\
WWW: \htmladdnormallink{\protect {\tt{http://www.slu.edu/colleges/AS/physics/profs/comer.html}}}
{http://www.slu.edu//colleges/AS/physics/profs/comer.html}\\
\bigskip
\hfill ISSN: 1527-3431

%\begin{figure}[ht!]
%\centering
%\includegraphics[width=160mm]{DGRAV_logo_horizontal.pdf}
%\end{figure}

\bigskip

DISCLAIMER: The opinions expressed in the articles of this newsletter represent
the views of the authors and are not necessarily the views of APS.
The articles in this newsletter are not peer reviewed.

\begin{rawhtml}
<P>
<BR><HR><P>
\end{rawhtml}
%{\bf \Large Contents:}
\end{flushleft}
\pagebreak
\section*{Editorial}

The next newsletter is due December 2017.  This and all subsequent
issues will be available on the web at
\htmladdnormallink 
{\protect {\tt {https://files.oakland.edu/users/garfinkl/web/mog/}}}
{https://files.oakland.edu/users/garfinkl/web/mog/} 
All issues before number {\bf 28} are available at
\htmladdnormallink {\protect {\tt {http://www.phys.lsu.edu/mog}}}
{http://www.phys.lsu.edu/mog}

Any ideas for topics
that should be covered by the newsletter should be emailed to me, or 
Greg Comer, or
the relevant correspondent.  Any comments/questions/complaints
about the newsletter should be emailed to me.

A hardcopy of the newsletter is distributed free of charge to the
members of the APS Division of Gravitational Physics upon request (the
default distribution form is via the web) to the secretary of the
Division.  It is considered a lack of etiquette to ask me to mail
you hard copies of the newsletter unless you have exhausted all your
resources to get your copy otherwise.

\hfill David Garfinkle 

\bigbreak

\vspace{-0.8cm}
\parskip=0pt
\section*{Correspondents of Matters of Gravity}
\begin{itemize}
\setlength{\itemsep}{-5pt}
\setlength{\parsep}{0pt}
\item Daniel Holz: Relativistic Astrophysics,
\item Bei-Lok Hu: Quantum Cosmology and Related Topics
\item Veronika Hubeny: String Theory
\item Pedro Marronetti: News from NSF
\item Luis Lehner: Numerical Relativity
\item Jim Isenberg: Mathematical Relativity
\item Katherine Freese: Cosmology
\item Lee Smolin: Quantum Gravity
\item Cliff Will: Confrontation of Theory with Experiment
\item Peter Bender: Space Experiments
\item Jens Gundlach: Laboratory Experiments
\item Warren Johnson: Resonant Mass Gravitational Wave Detectors
\item David Shoemaker: LIGO 
\item Stan Whitcomb: Gravitational Wave detection
\item Peter Saulson and Jorge Pullin: former editors, correspondents at large.
\end{itemize}
\section*{Division of Gravitational Physics (DGRAV) Authorities}
Chair: Peter Shawhan; Chair-Elect: 
Emanuele Berti; Vice-Chair: Gary Horowitz. 
Secretary-Treasurer: Geoffrey Lovelace; Past Chair:  Laura Cadonati; Councilor: Beverly Berger
Members-at-large:
Duncan Brown, Michele Vallisneri, Kelly Holley-Bockelmann, Leo Stein, Lisa Barsotti, Theodore Jacobson.
Student Members: Megan Jones, Cody Messick.
\parskip=10pt

\vfill\eject

\section*{\centerline
{we hear that \dots}}
\addtocontents{toc}{\protect\medskip}
\addtocontents{toc}{\bf DGRAV News:}
\addcontentsline{toc}{subsubsection}{
\it we hear that \dots , by David Garfinkle}
\parskip=3pt
\begin{center}
David Garfinkle, Oakland University
\htmladdnormallink{garfinkl-at-oakland.edu}
{mailto:garfinkl@oakland.edu}
\end{center}

Rainer Weiss, Kip Thorne, Barry Barish and the LIGO Scientifc Collaboration have been awarded the 2017 Princess of Asturias Award for Technical and Scientific Research.

The LIGO team has been awarded the 2017 Group Achievement Award of The Royal Astronomical Society.

Gabriela Gonz\'alez and the LIGO Scientific Collaboration have been awarded the Bruno Rossi prize of the High Energy Astrophysics Division of the American Astronomical Society.

Barry Barish and Stan Whitcomb have been awarded the Henry Draper Medal of The National Academy of Sciences.

David Reitze, Gabriela Gonz\'alez, and Peter Saulson have been awarded The National Academy of Sciences Award for Scientific Discovery.

Gabriela Gonz\'alez has been elected to The National Academy of Sciences and to The American Academy of Arts and Sciences.

Nergis Mavalvala has been elected to The National Academy of Sciences and to The American Academy of Arts and Sciences.

Gabriela Gonz\'alez has been named by the journal {\it Nature} to their 2016 list of {\it Ten people who mattered this year.}

David Shoemaker has been elected Spokesperson of the LIGO Scientific Collaboration.

Hearty Congratulations!

\vfill\eject
\section*{\centerline
{DGRAV student travel grants}
}
\addtocontents{toc}{\protect\medskip}
\addcontentsline{toc}{subsubsection}{
\it DGRAV student travel grants, by Beverly Berger}
\parskip=3pt
\begin{center}
Beverly Berger, LIGO
\htmladdnormallink{beverlyberger-at-me.com}
{mailto:beverlyberger@me.com}
\end{center}

SUPPORT DGRAV STUDENT TRAVEL GRANTS
The American Physical Society's Division of Gravitational Physics (DGRAV) has launched a campaign to endow the DGRAV Student Travel Grants.  We invite you to help us ensure its success!  
First awarded in 1999 by DGRAV’s predecessor, the Topical Group in Gravitation, these grants provide partial travel support to allow DGRAV student members to present their work at the APS April Meeting.  They not only benefit the students, but also provide other participants the opportunity to meet the grant recipients and learn about their work. 
Over the past few years, support for student travel has become one of DGRAV's core missions and, accordingly, consumes a major fraction of the DGRAV budget.  
As the field, the number of students in the field, and the significance of the APS April Meeting to the field have increased, so have the applications for travel grants.  
For example, the number of applicants jumped from 22 
in 2014 to 45 in 2015-the General Relativity Centennial.  While the 2016 number was ``only'' 28, this level of demand is still unsustainable.  As a consequence, DGRAV can no longer support all the high quality applications we receive.  Therefore, the purpose of this endowment is to provide funding for this activity at a level beyond that achievable now by DGRAV.  Our campaign goal is to raise \$75,000, enough to support 10 students every year alongside those funded from the DGRAV operating budget.  We hope that future growth in the endowment will match growth in the demand.  At the time of writing (6/17/2017) we have raised \$46,250 towards our goal.  We need {\emph {your}} help to get all the way there.  

More information is available at \hfill \\
\htmladdnormallink
{\protect {\tt{https://www.aps.org/about/support/campaigns/dgrav/index.cfm}}}
{https://www.aps.org/about/support/campaigns/dgrav/index.cfm}
including instructions on how to donate.  Donors will be recognized on the campaign website, in the DGRAV newsletter {\it Matters of Gravity}, as well as on the DGRAV website (unless you prefer that your gift not be recognized publicly). 

Contact information:
Beverly K. Berger, Chair, DGRAV Student Travel Grants Endowment Campaign, \htmladdnormallink{\protect {\tt{beverlyberger@me.com}}}
{mailto:beverlyberger@me.com}
Irene Lukoff, APS Director of Development, \htmladdnormallink{\protect {\tt{lukoff@aps.org}}}
{mailto:lukoff@aps.org}
Tora Buttaro, APS Donor Relations Program Manager, \htmladdnormallink{\protect {\tt{buttaro@aps.org}}}
{mailto:buttaro@aps.org}
\vfill\eject

\section*{\centerline
{The Discovery of GW170104}}
\addtocontents{toc}{\protect\medskip}
\addtocontents{toc}{\bf Research Briefs:}
\addcontentsline{toc}{subsubsection}{
\it  The Discovery of GW170104, by Jenne Driggers and Salvatore Vitale}
\parskip=3pt
\begin{center}
Jenne Driggers, LIGO Hanford Observatory 
\htmladdnormallink{jenne.driggers-at-ligo.org}
{mailto:jenne.driggres@ligo.org}
\end{center}
\begin{center}
Salvatore Vitale, MIT 
\htmladdnormallink{salvatore.vitale-at-ligo.mit.edu}
{mailto:salvatore.vitale@ligo.mit.edu}
\end{center}

LIGO  has detected gravitational waves for the third time (or fourth, if you
believe that LVT151012 is of astrophysical origin), on January
4, 2017.

The source, a binary black hole coalescence dubbed GW170104, was
discovered with the two NSF-funded Advanced LIGO detectors just over a month after the beginning of the second
observing run, which started on November 30, 2016. The LIGO Scientific Collaboration, with the Virgo Collaboration, publicly announced the findings on June 1, 2017. 

A year-long break between the first and second observing runs was
spent implementing a series of upgrades at both sites.
%runThe LIGO detectors underwent a  series of upgrades between the first
%and second observing runs; focussing on improving the low-frequency
%sensitivity at the Livingston observatory, as well as increasing the duty
%cycle at both sites.  During this
%time of upgrades, t
The Hanford observatory focused on increasing the
laser power circulating in the interferometer.  While we are able to
operate at double the first run's input power, noise performance
challenges encouraged us to back down to only 50\% more input power.
This work was still a success in that we have developed infrastructure
to damp unstable opto-mechanical modes which can cause the
interferometers to fall out of resonance, and diagnosed many other
technical issues that will become important when we do go to higher
power after this run.  The Hanford site also made great strides in
increasing robustness against various environmental conditions,
particularly times of high wind.  Ground tilt sensors now measure how
much the floor below the interferometer tilts, and once we take that
into account we are able to achieve roughly 80\% duty cycle, including
one continuous data stretch over 71 hours long. At the same time, the
Livingston observatory found some sources of scattered light inside
the vacuum system, and were able to improve their low frequency
sensitivity between 50\,Hz and 200\,Hz. 

%With the Advanced LIGO interferometers operating so reliably 
%through both our first and
%second observing runs, we have learned a great deal about binary black
%hole systems.   
The hard work paid off shortly after the detectors came back online, with the discovery of GW170104.
%GW170104, was once again a binary black
%hole coalescence. 

%GW170104 was discovered just over a month after the beginning of the
%second observing run, started on November 30, 2016. 
%
But not everything
went exactly as planned. Under normal circumstances, low-latency algorithms
searching for compact binaries in LIGO data would alert the
observatories and data analysts when a
significant trigger is found in both data streams. However, at the time of the
event, the Hanford site suffered a problem with the subsystem that reports on the
status of the calibration. The signal was thus first identified by manually
inspecting low-latency triggers in the Livingston data only. Shortly after, it
was verified that the calibration at both sites was in a nominal state, and the
data from Hanford included in the analysis.

It was worth the struggle, since very soon it became clear that
GW170104 was every bit as interesting as the two clear detections made in the
first observing run.

A preliminary map with the source localization was released to partner
astronomers, and followed up by over 30 facilities for potential
electromagnetic or neutrino counterparts. At the time this newsletter is written, no
significant counterparts have been claimed.

With a false alarm ratio smaller than 1 in 70,000 years of coincident
data, and a probability of being of astrophysical origin which differs from 1 by
3\,$\times$\,$10^{-5}$, GW170104 stands clearly above the instrumental
background.

Among the many reasons why LIGO's first discovery, GW150914, was
extraordinary is that it showed that stellar mass black holes can have
large masses. Stellar mass black holes found in X-ray binaries all have masses
smaller than 15~M$_\odot$. The component black holes of GW150914 were
a factor of two heavier, 
suggesting their progenitor stars formed in a low metallicity 
environment. The second confident event, GW151226, was made of smaller black holes, of
masses compatible with what is found in X-ray binaries. GW170104 is in the
middle, with a primary object at 31~M$_\odot$ and secondary at 19~M$_\odot$. 

LIGO is thus accessing a population of black holes covering a large
range of masses. Assuming that the mass function of black holes in binaries
follows a power law, we can measure its slope $\alpha$. Using the 4 
detections we have thus far, we have obtained $\alpha = 2.3^{1.3}_{-1.4}$. The
median of our measurement is remarkably close to the exponent of the 
Salpeter mass function. More detections will allow us to further improve this
measurement. 

The detection of GW170104 has allowed us to slightly tighten our
measurement of the merger rate for binary black holes. The discoveries of
the first observing run had led to an estimate of 9 -- 240\,Gpc$^{-3}$
yr$^{-1}$. Including GW170104, we obtain 12 -- 213\,Gpc$^{-3}$
yr$^{-1}$. This confirms that black hole mergers are common in the universe, 
and will be detected in large numbers in the future.

One of the unsolved problems about compact binaries such as those detected
by LIGO is how and where they formed. Two scenarios are typically
considered: isolated binary evolution and dynamical formation. In the
first case the formation happens in galactic fields, via a common
envelope evolution or other mechanisms. In the second case, the two
black holes dynamically form a bound system in a dense environment
such as a globular cluster, or near a galactic nucleus. Whether both
channels happen, and with which relative frequency, is an open
question in astrophysics. 

Black hole spins are one of the keys to answer these
questions. Whereas dynamical formation quite naturally results in
random 
spin orientations, isolated binary evolution typically predicts
binaries with only moderate spin misalignment with respect to the orbital
angular momentum. This implies that the mass-weighted projection of
the total spin along the orbital angular momentum, also known as
effective spin, can have both signs for systems evolved dynamically,
while it should be positive for isolated binaries. 

For GW150914 the posterior of the effective spin was centered at zero,
whereas for GW151226 it preferred positive values, which is compatible
with both formation channels. The situation is different for GW170104,
for which roughly 80\% of the posterior probability is for negative
effective spin. While positive values are not excluded, there seems to
be a preference for a dynamical formation. The evidence is not
conclusive, but it shows what kind of information is accessible with
gravitational waves. As more sources are detected, we will be able to further
study the formation channels of compact binaries.

LIGO's detections give us access to regions of the space time where
the gravitational field is strong and dynamical. As such, they are the
ideal tool to stress test general relativity. Over the last several decades, a
large number of alternative theories of gravity have been proposed. In
some of them, one of the fundamental laws of physics, Lorentz
invariance, is broken, which affects the dispersion relation for
gravitational waves. In these theories, gravitational waves do not
travel at the speed of light but rather have a frequency-dependent
group velocity.  

We have used GW170104, as well as the events detected in the first
observing run, to put upper limits on the magnitude of Lorentz
violation tolerated by our data. The bounds we obtain are weaker than
those existing from electromagnetic or neutrino observations. However,
they are still important! In fact, there are theories in which Lorentz
invariance is only broken in the gravity sector, leaving the neutrino
and electromagnetic sectors unaffected. Ours are the first constraints
 based on gravitational waves, and the first tests of superluminal propagation in the gravitational sector.

We have also considered a modification of general relativity in which
gravitons are dispersed in vacuum like massive particles. This way, we
obtained an upper bound for the graviton mass at
7.7\,$\times$\,10$^{-23}$\,eV/c$^2$,  
improving the results of the first observing run. 

Finally, we performed unmodeled tests. First, we augmented the
waveform model with extra phase coefficients not normally present in
general relativity, and we measured them. For all these extra
parameters, the posterior distributions contain zero, which is the
general relativity value. Next, we verified that the early portion of
the detected signal yield mass and spin estimates which are consistent
with what can be measured from the final portion of the waveform. 
In this case too general relativity is sufficient to describe
GW170104. While we are thrilled that Einstein's gravity has brought us
so far, allowing us to successfully detect and characterize GW170104
and the previous sources, we will keep testing.

The second observing run will last for another few months, probably
until the end of the summer. It will not be business as usual,
though. The Virgo interferometer in Italy has successfully locked
and is collecting data, although still at a low sensitivity. As Virgo
improves, a short joint observing run could happen by the end of the
summer. A third detector could significantly reduce the uncertainty in
the sky position of the sources, increasing the chances of successfully
detecting an eventual electromagnetic counterpart to
gravitational-wave detections. Work has already begun in the LVC to be
sure everything is in place for when Virgo data will be used for
joint analysis.

Once the second observing run is complete, both observatories will
again go offline, likely for more than one year, for a large set of
upgrades and some vacuum facilities maintenance.  Many of these
upgrades to the interferometers are based on lessons learned over the
past two years.  The lasers at both sites will be replaced which will
allow us to inject more laser power into the interferometer with less
beam motion jitter than our current high power lasers.  The Livingston
observatory, and perhaps also Hanford, will inject squeezed quantum
states of light into the interferometers.  The goal is to have 3\,dB
of effective squeezing, which is equivalent to again doubling the
input laser power, but without the challenges of the thermal input of
higher power.  Some of our other upgrades include installing baffles
to catch many more stray light beams, and prevent them from
potentially scattering back into the main laser beam path.  The
so-called reaction masses that sit about 5\,mm behind the highly
reflective end mirrors will be replaced with annular versions rather
than the current disc geometry.  This will allow us to continue using
them as a quiet actuation reference for the test mass mirrors without
suffering from thin film gas damping due to molecules bouncing between
the surfaces of the mirror and the reaction mass, and potentially
reducing the effect of electromagnetic charge on our calibration.  Our
readiness to begin our third observing run will be driven by achieving
a significant increase in sensitivity over the current run.  

While 2015 has been the year of the discovery, 2017 is truly the year
of the beginning of gravitational-wave astronomy. Already after three
confident detections we have significantly broadened our understanding
of black holes, including their masses and spins. We have put general
relativity under scrutiny, and found it consistent with the detected
signals.

As more and different types of sources are discovered,
information from gravitational waves have the potential to
dramatically change what we
know about compact objects, matter, and gravity. 

Fasten your seatbelts.

\vfill\eject

\section*{\centerline
{Remembering Vishu}}
\addtocontents{toc}{\protect\medskip}
\addtocontents{toc}{\bf Obituary:}
\addcontentsline{toc}{subsubsection}{
\it  Remembering Vishu, by Naresh Dadhich and Bala Iyer}
\parskip=3pt
\begin{center}
Naresh Dadhich, ICUAA, Pune 
\htmladdnormallink{nkd-at-iucaa.in}
{mailto:nkd@iucaa.in}
\end{center}
\begin{center}
Bala Iyer, Tata Institute of Fundamental Research 
\htmladdnormallink{bala.iyer-at-icts.res.in}
{mailto:bala.iyer@icts.res.in}
\end{center}

C V Vishveshwara, or   Vishu, is associated  in the minds of  most of us with  quasi-normal modes or the ringdown of a black hole.  The prediction that his simple calculations made was dramatically verified after 46 years with the discovery of gravitational waves by  LIGO. It was almost a year before he breathed his last on 16 Jan 2017 in Bengaluru. It was, therefore, most fortituous that he could experience exhilaration and satisfaction of his contribution when the whole world was cheering and applauding. The black hole man of India  will be  remembered for a long time not only for his seminal contributions to understanding black holes but fondly for  the word pictures and the Sydney Harris like cartoons he created to  share with  his professional colleagues and the lay public the  esoteric consequences of Einstein's general theory of relativity. His talks inspired generations of students to a career in science and via  the activities at the Jawaharlal Nehru Planetarium and Bangalore Association for Science Education the inspiration lives on. 

Vishveshwara was born on March 6, 1938 in Bangalore. He had his schooling there and then went on to Mysore University for further studies. He obtained the B.Sc.(Hons) degree in 1958 and the M.Sc. Degree from Central College of the then Mysore university in 1959. He then went to USA  for higher studies. After getting his A.M. from Columbia University, New York, in 1964 he moved to University of Maryland from where he got his Ph.D. in 1968.  His thesis advisor was Prof. C.W. Misner, the M of the directory of the universe, MTW.  His thesis subject was "Stability of Schwarzschild Metric". After stints as a post doctoral fellow and  a visiting faculty member, at  Institute of Space Studies (1968-69), Boston University (1969-72) , New York University (1972-74), University  of Pittsburgh (1974-76), Vishu  returned to Bangalore in 1976 and joined the Raman Research Institute. He moved from there, in December 1992, to the Indian Institute of Astrophysics, Bangalore as a Senior Professor, from where he retired in 2005.

One of the most important and bizarre predictions of General Relativity is the existence of black holes – objects from which nothing can come out including light. It marks a surface which can only be crossed one way but not the other: things can fall in but nothing can come out.  A brief historical aside is not out of place to give a flavour of the times when Vishu's important papers were written.

Relativity  revolutionized our understanding of space and time by first uniting them  into a flat four dimensional  space-time  in special relativity and subsequently for describing gravity making it curved and dynamic in General Relativity.   Gravity is  no longer an external force but synonymous with  the  geometry of space-time.  In  1915, Einstein finally arrived at the  correct field equations completing the quest he began in 1907 to obtain General Relativity, his relativistic theory of gravitation. Mathematically the equations were complicated and so he was surprised that within a year Karl Schwarzscild discovered an exact solution of these equations representing a spherically symmetric, asympotically flat, vacuum solution, whose outer region is strictly static. The solution had an unusual feature that a certain component of the metric vanished while another diverged  at what was referred to as  the Schwarzschild singularity or better the Schwarzschild surface. Though in 1939 Oppenheimer and Snyder showed that a person who rides through this surface on  an imploding star will feel no infinite gravity or see no breakdown of physics there, these results were not taken seriously due to the mental connotation associated with the word `singularity' and due to the simple dust model used in the treatment. These objects  were referred to as frozen stars in the Soviet Union and collapsed stars in the west. The realization that this was due to a choice of coordinates or a coordinate singularity was a long time coming and conclusively settled in 1958 by Finkelstein (and later in 1960 by Kruskal) who discovered a new reference frame for the Schwarzschild geometry. In December 1967, in  his lecture  on ``Our universe, the known and unknown'', John Wheeler christened these objects Black Holes, an idea  that intrigues and fascinates scientists and the lay public even to this day.

General Relativity  is a complex mathematical theory and often  involves subtleties in its physical interpretation related to the choice of coordinates used in its formulation. Can one use a description  using  more well-behaved coordinates? Even if mathematically a black hole solution exists, the possibility of it being a physical object in nature depends on whether it is stable. If the   black hole is an object from which no information can escape, how can one look for it?  Can one provide a mathematically  elegant description of the physical effects of a rotating black hole  like gyroscopic precession? Vishu's seminal research centered on these topics and earned him the fond title of  Black Holy man of India!

Among Vishu's classics on this topic is a brief elegant  paper using Killing vectors to provide a coordinate invariant distinction between the stationary Kerr and static Schwarzschild black hole cases and the consequent  existence of the ergosphere [1]. Regarding this work  Jacob Bekenstein commented [2]:  ``I was familiar with the Vishu theorem that the infinite redshift surface of a static black hole is always the horizon. At that time black hole physics was just getting started and such neat relations between black hole features were rare. Vishu's theorem was a welcome hard fact in the middle of such folklore and helped clarify in mind what black holes were about.  At the conference (GR6) I had a long talk with him and I vividly remember being impressed by the range of research problems he had going simultaneously.''

Vishu was the first to prove the stability of non-rotating black holes under linear perturbations [3]. Regarding this  Brandon Carter remarked [2]: ``Vishu  was one of the first to appreciate the importance of this problem and who played an important role in persuading others to take the problem seriously as something of  potential astrophysical relevance by providing the first convincing proof that at least in one case namely the Schwarzschild solution, such an equilibrium state can be stable.'' Elaborating further Bernard Whiting wrote [2]: ``Vishveshwara's original discussion of stability showed that there was no superficial case establishing the instability basically by dealing with single modes and by demonstrating the positivity of effective potentials. Establishing pointwise boundedness requires use of more refined tools leading to a method that differs markedly in substance but not at all in essence from the relatively simple positive potential approach. Vishu made a number of significant breakthroughs...''  

Vishu was the pioneer who  explored how black holes respond when externally perturbed [4] and proved  that regardless of  the perturbation, Schwarzschild black holes get rid of any deformation imparted to them by radiating gravitational waves with a frequency and decay time that depended only on their mass. These characteristic waves are technically termed quasi-normal modes, which is why after the announcement of the gravitational wave detection by LIGO Vishu laid the claim to  the nom de plume  ``Quasimodo of black holes''.  Quasi-normal modes are like  the dying tones of a bell struck with a hammer and are referred to as  the ringdown radiation. Vishu's work is fundamental to our understanding of black holes and began a new chapter in how to study them.

Many of us met Vishu during  the Einstein Centenary symposium at Physical Research Laboratory, Ahmedabad in 1979. Though we have other  wonderful memories of the symposium the most memorable one was Vishu's lecture entitled `Black Holes for Bedtime'. It was a magical experience; an exotic cocktail of science, art, humour and caricature. Equations  were not necessarily abstract and unspeakable and could  well be translated in the best literary tradition if you were Vishu! 

At Raman Research Institute and later Indian Institute of Astrophysics  Vishu explored problems in classical general relativity with possible astrophysical implications. Perturbations of black holes in general relativity  carry signatures of the effective potential  around them and one could  look for them by examining neutrinos in gravitational collapse or ultracompact objects. Could one discern possible differences between  black hole solutions in general relativity and other theories of gravity by looking at their quasi-normal modes and the properties of their horizons?  How different are  black hole solutions in cosmological backgrounds from those in the usual asymptotically flat ones?  How does one use the  Frenet-Serret formalism to study gyroscopic precession, general relativity analogs of inertial forces, and characterize black holes in higher dimensions in a covariant and geometric manner?  Other  mathematical issues studied related to separability of different spin perturbations  in general relativity, the role of the Killing tensor in separability of wave equations among others. It was always  a pleasure working with Vishu. There was no pressure, no generation gap, a natural possibility to grow and contribute your best, an easy personal rapport, a refreshing sense of humour, an unassuming erudition and most importantly a warm and wonderful human being.

Together with J.V. Narlikar, Vishu played a key role in bringing long  due recognition to the doyens of general relativity P.C. Vaidya and A.K. Raychauduri.  A volume entitled {\it Random walk in relativity and cosmology} co-edited by them was released in 1986 at RRI and its royalties supplemented by royalties of the International Conference on Gravitation and Cosmology (ICGC) proceedings were used to set up the Vaidya-Raychaudhuri endowment lecture of the Indian Association for General Relativity and Gravitation (IAGRG). Vishu was closely involved in the  group that initiated, planned and  organized  UGC Schools on general relativity  and cosmology in the 1980’s. The motivation was to extend Indian research in exact solutions in  general relativity to modern research frontiers in cosmology, early universe and relativistic astrophysics. This led to the ICGC meetings organized every four years because it was recognized that due to limited resources, participation of the Indian researchers in the International Society of General Relativity and Gravitation (ISGRG) meetings was very limited.  Creating an opportunity for the  IAGRG community to  interact with international experts on front line research areas in relativity and cosmology in India was needed to  assist in improving the quality and relevance of general relativity research in India. These meetings also brought out   the cartoonist in Vishu during the first ICGC in Goa.  Between sessions cartoons would appear on the screen anonymously and by the end of the meeting there were multiple reprint requests for them! Staid Cambridge University Press was happy to include them in the proceedings and   Vishu's  cartoons in the ICGC proceedings were a treat to look forward to. The  series of cartoons on gravitational waves in those proceedings deserves special mention.  Alas they are incomplete since he could not make one after the discovery.  Just on the day he passed away Nils Andersson wrote Vishu an email:  ``I have recently done something that I think might amuse you. I have written a little book involving Einstein, relativity and a fair bit of fictional freedom. Now, I think it is fair to say that my attitude to this project has been heavily inspired by your story-telling, your drawings and the bathtub book [5].''

Vishu's  public lectures  inspired a number of students all over the country.   His lectures at Bangalore Science Forum, started by his Guru Dr H. Narsimiah, always drew huge numbers. He was a best-seller. And, he never disappointed the audience. Without diluting the profound ideas that he would discuss, he would lace the talks with subtle humour that came seamlessly.  At Vishu's passing, countless echoed Sathyaprakash who  exclaimed ``This is devastating. I have lost a teacher, a mentor and a friend. More than anything else we are going to miss his `serious' sense of humour in all walks of life, especially science.''

Together with a committed group that included Sanjay Biswas, Vishu was involved in bringing out Bulletin Of Sciences from 1983-1993 to set up a forum  to seriously  address the social impact of science and technology. To find the means of sustaining it financially he co-edited with Sanjay Biswas and D.C.V. Mallik an interesting volume called {\it Cosmic Perspectives} that was dedicated to the memory of M. Vainu Bappu. Together with A. Ratnakar Vishu was instrumental in setting up the RRI Film Club in the 1980’s to get access to movie classics from National Film Archives in Pune and from the consulates like the German and French ones.

Jawaharlal Nehru Planetarium (JNP), Bangalore  is a wonderful testament  to  Vishu’s vision which  showcases his multi-faceted personality in science communication and education. Starting as its founder director in 1988, Vishu brought together a dedicated and talented team and inspired them to build a  world class planetarium scripting  unique shows integrating the best in science and astronomy with the best  in  world and Indian history, art, literature and music. By example he set up high standards for all the JNP personnel and mentored them  till the very end. But JNP was not to be just a theatre.  It had to  play a role in science education in the city. Thus  in 1992 Bangalore Association for Science Education (BASE)  was set up by Vishu to  systematically expose, attract  and mentor students from elementary school, high school and colleges for a career in science. It may surprise many that in spite of being a pure theorist, Vishu firmly believed in doing science experiments. Via activities like `Science in Action' he  emphasized the importance of bringing out in young students the joy of seeing scientific phenomena. That was a way to attract them to science. In fact this philosophy of `doing' science underlined every activity that was visualized at JNP in the coming years. SEED (Science Education in Early Development) for middle school children, SOW (Science Over the Weekends) for high school children and at the pinnacle of the  educational programs, REAP (Research Education Advancement Programme) for undergraduate students. SEED, SOW and REAP, all have a very strong presence of experiments that make the programs dynamic and vibrant and endearing to students. During the last twenty years, all these programs have seen a steady growth in number of students attending them and also in attracting quality students with a potential to excel in a career in science.  No wonder that more than hundred students who passed through JNP are either pursuing PhD programs or have completed them. Some of them are faculty at  institutions such as ICTS, JNCASR and IMSc.  Finally, setting up of a science park at JNP was also his initiative. In the original plan drawn up in 1997, an `Antigravity Cottage' that mimics the famous `Mystery Spot' in the US and some other places had been envisaged. It was realised in 2016.

When the gravitational wave  discovery by LIGO was announced last year, Vishu was elated. We have never seen him so high, thrilled by the possibility that soon there would be events where the quasi-normal modes would be even more strong. The profoundness of this discovery is in the realization that the black hole, which is purely a geometric object without any hard surface boundary rings under perturbations like  a material object. It is indeed the most telling and `visible' defining property of a black hole. And Vishu was its discoverer.  By all accounts, it is a discovery that will endure in relativity textbooks. By that benchmark, there are only a few other contributions from India  like the Raychaudhuri equation and Vaidya's radiating star  that will make the grade. On the other hand this discovery sits alongside the celebrated result that a black hole has no hair –the  `No Hair' theorem.  Most important of all,  it is one of the few predictions that have been brilliantly verified by the observation of gravitational waves produced by the merger of two black holes. The observed profile has very uncanny resemblance with what Vishu had plotted long ago back in 1970. There are very few  predictions which are actually verified by experiment and observation. Vishu's black hole ringdown is one among those few. This is the true and ultimate measure of a seminal insight.

We will miss you Vishu even as we try very  hard  to follow your favorite lines from Machado: ``Traveller there is no Path, Paths are made by Walking.''

Vishu is survived by his wife Saraswati and two daughters Smitha and Namitha.

[1] Generalization of the "Schwarzschild Surface" to Arbitrary Static
and Stationary Metrics, C. V. Vishveshwara, J Math. Phys., 9,
1319 (1968).

[2] Black Holes, Gravitational Waves and the Universe, Essays in honor of C.V. Vishveshwara, Eds. B. R. Iyer and B. Bhawal, Kluwer, (1999).

[3] Stability of the Schwarzschild Metric, C. V. Vishveshwara, Phys.
Rev. D, 1, 2870 (1970),
 
[4] Scattering of Gravitational Radiation by a Schwarzschild BlackHole,
C. V. Vishveshwara, Nature, 227, 936 (1970)

[5] Einstein's Enigma or Black Holes in My Bubble Bath, C.V. Vishveshwara, Springer-Verlag, Berlin-Heidelberg (2006).

Reproduced with permission from:
CURRENT SCIENCE (Vol. No. 112, 25 February 2017, pp. 866-868).
\vfill\eject

\section*{\centerline
{Remembering Cecile DeWitt-Morette}}
\addtocontents{toc}{\protect\medskip}
\addcontentsline{toc}{subsubsection}{
\it  Remembering Cecile DeWitt-Morette, by Yvonne Choquet-Bruhat}
\parskip=3pt
\begin{center}
Yvonne Choquet-Bruhat, IHES Paris 
\htmladdnormallink{ycb-at-ihes.fr}
{mailto:ycb@ihes.fr}
\end{center}

(Editor's note: Cecile DeWitt-Morette passed away on May 8, 2017. Below is a translation of some remarks given by Yvonne Choquet-Bruhat during the ceremony held at IHES in which Cecile was elevated to the rank of officer of the L\'egion d'Honneur in 2011.  The English translation of this text was kindly provided by IHES which is preparing an obituary for Cecile for its next newsletter.  There will be an additional obituary for Cecile in the December 2017 issue of Matters of Gravity). 

I am happy and proud that Cecile has chosen me to present to her the
insignia of an officer of the L\'egion d'Honneur. France certainly owed Cecile this decoration because she is largely responsible for the renaissance and flourishing, starting in the 1950s, of French theoretical physics, a discipline that had sadly declined because of
both the war and the dominating influence of an aging De Broglie.

As everyone knows, Cecile set up the Les Houches Summer School as a compensation for having married a foreigner, Bryce DeWitt. Cecile was able to build this school and run it thanks to her energy, her determination and her understanding of human relations, among others. She led the school for over twenty years and it is still going strong today. Not only did she rescue French theoretical physics from its moribund state, she also trained the greatest physicists around the world: her astuteness in choosing topics and speakers, her clear-sightedness, her courage and tenacity in organizing work and daily living are well known by all physicists and I need not dwell on this subject any further. I will only add here the role Cecile played in the creation of the wonderful Institute that is IHES, founded by the businessman and mathematician, Motchane.

It so happened that Cecile and I were in the same year in fifth grade at the Lyc\'ee Victor Duruy in Paris, but in separate sections. I only met Cecile twenty years later, when I went to see her at the Institut Poincar\'e, on the advice of Georges Darmois, to ask to take part in her summer school for two weeks. The handsome young woman replied very firmly: ``the school is all (two months at the time) or nothing.'' It was to be nothing.

I only really got to know Cecile at the conference she had organized in 1957 in Chapel Hill, the first one ever held by the General Relativity and Gravitation (GRG) Society. Its founding members were the great relativists of the time: J.A. Wheeler, A. Lichnerowicz, M.A. Tonnelat, P. Bergmann, V.A. Fock, \dots and of course, Bryce DeWitt. Bryce was a great physicist and a man who loved adventure, one worthy of Cecile, although perhaps not always easy to live with, but Cecile also knew how to manage family life - a husband and four daughters - which I also admired. The Chapel Hill conference was, like everything else Cecile does, a great success. It marked the advent of General Relativity as a true physics theory, based on rapidly developing mathematics: analysis on manifolds and non-linear partial differential equations. The GRG Society, created at the instigation of the DeWitts, now has over one thousand members and its triennial conferences are held across the globe, attracting ever increasing numbers of participants.

Having got to appreciate Cecile in Chapel Hill, I invited her to Paris. She gave a series of conferences at the Institut Poincar\'e on her favorite subject, the Feynman path integral, which fascinated the audience, myself included. Unfortunately, the difficulties of interpreting the Feynman path integral in terms of standard rigorous mathematics were discouraging to someone like me with a more down-to-earth mindset. In contrast, Cecile worked on what was then modern differential geometry and even offered to translate into English a fairly elementary book I had written on the subject. I did not think it worthy of her and I suggested instead that we write together a more comprehensive book that would also contain analysis and applications to physics, which could be used by physicists. That was how we came to write {\it Analysis, Manifolds and Physics} and then a second volume, {\it 92 Applications}, which became Volume II in a subsequent, expanded edition. It was a great pleasure to write these books with Cecile. She is like me, even more so, in always wanting to learn new things. She is a great worker who likes to take every task to completion, every calculation in complete detail, and every theory in full generality.

But Cecile is also open to suggestions and her friendliness makes it very pleasurable to work with her. Naturally, her remarkable interpersonal and organizational skills also contributed to the completion of our books and their publication, just like they contributed to everything else she undertakes. Unfortunately for me, our different specialties brought an end to our writing together but happily, not to our friendship. Cecile had never given up her research on Feynman's path integral, which she had started after meeting Dyson and Feynman in the United States; the doctoral thesis that she prepared in Dublin then defended in Paris was on a completely different topic. Cecile, together with her many students, obtained deep and varied results on this beautiful and mysterious integral. A number of them can be found in a book written with Pierre Cartier on functional integration, published by Cambridge University Press. After the passing of Bryce, Cecile also recently had a book published by Springer: {\it The pursuit of Quantum Gravity, memoirs of Bryce DeWitt.}

Dear Cecile, it is with my warmest congratulations that I give you the insignia of officer of the L\'egion d'Honneur.

\vfill\eject
\section*{\centerline
{Remembering Larry Shepley}}
\addtocontents{toc}{\protect\medskip}
\addcontentsline{toc}{subsubsection}{
\it  Remembering Larry Shepley, by Richard Matzner and Mel Oakes}
\parskip=3pt
\begin{center}
Richard Matzner, University of Texas at Austin 
\htmladdnormallink{matzner2-at-physics.utexas.edu}
{mailto:matzner2@physics.utexas.edu}
\\Melvin Oakes, University of Texas at Austin 
\htmladdnormallink{oakes-at-physics.utexas.edu}
{mailto:oakes@physics.utexas.edu}
\end{center}

Lawrence Charles ``Larry" Shepley was born August 11, 1939, in Washington D.C., to Jack Mandel and Ida Bernstein Shepley. Larry's father was a  licensed civil, electrical, and structural engineer, who worked for the Federal Power Commission where he participated in the licensing of many of the large dams on US navigable rivers. Larry attended Woodrow Wilson High School in Washington DC, where he excelled in science and mathematics. Larry graduated from Swarthmore College with a bachelor's degree in physics in 1961. He earned a master's degree and a doctorate in physics in 1963 and 1965 from Princeton University, where he studied under John Archibald Wheeler and Charles W. Misner. His dissertation was entitled ``SO(3, R)-Homogeneous Cosmologies." Following a two-year post-doctoral fellowship at the University of California, Berkeley, under Abraham H. Taub, Larry joined the Physics Department at the University of Texas at Austin in 1967 as an Assistant Professor of Physics, and was in 1970 promoted to Associate Professor. 

\noindent He served the UT Physics Department in several capacities, as Vice-Chair for Graduate Affairs and Graduate Adviser, as Associate Director of the Center for Relativity, as Departmental Minority Liaison Officer, as the Chair of the Teaching Assistant's Committee, and in other advisory capacities. He was the Chair of the Equal Opportunities Committee of the College of Natural Sciences at UT Austin. He taught classes at all levels, from basic freshman physics for non-science students to specialized graduate courses. 

\noindent In addition, Professor Shepley served as a member of the organizing committee of the Texas Symposia on Relativistic Astrophysics, as an instructor at the Curso Centroamericano y del Caribe de Fisica, and as a member of the American Physical Society Committee on Civil Defense. He authored or co-authored/edited over 50 scholarly articles and four books. 

\noindent Professor Shepley retired in 1995 but continued to be active.  He was the Chair of the Texas Section of the American Physical Society and a State Contest Director of the University Interscholastic League. 

\noindent Larry was a longtime member of the Gilbert and Sullivan Society of Austin, serving as a board member and president.  Larry loved to travel, visiting all continents, including Antarctica. He died of congestive heart failure on December 30, 2016. He is survived by his sister, Lona Piatigorsky, her husband, Joram, their two sons and their grandchildren, and by many friends. At his request, there were no memorial services or ceremonies and no solicitations in his memory. Larry's final sentiments, in his own style, were, {\it ``Farewell and Good Luck."}

\vfill\eject
\section*{\centerline
{Remembering Marcus Ansorg}}
\addtocontents{toc}{\protect\medskip}
\addcontentsline{toc}{subsubsection}{
\it  Remembering Marcus Ansorg, by Bernd Br\"ugmann and Reinhard Meinel}
\parskip=3pt
\begin{center}
Bernd Br\"ugmann, Friedrich-Schiller-Universit\"at Jena 
\htmladdnormallink{bernd.bruegmann-at-uni-jena.de}
{mailto:bernd.bruegmann@uni-jena.de}
Reinhard Meinel, Friedrich-Schiller-Universit\"at Jena 
\htmladdnormallink{R.Meinel-at-tpi.uni-jena.de}
{mailto:R.Meinel@tpi.uni-jena.de}
\end{center}

Additional contributors: Bruno Giacomazzo, Dorota Gondek-Rosinska, Eric Gourgoulhon, J\"org Hennig, Jose Luis Jaramillo, Rodrigo Panosso Macedo, Gernot Neugebauer, David Petroff, Luciano Rezzolla, Nikolaos Stergioulas, Loic Villain
 
On December 2nd 2016, our close friend and collaborator Marcus Ansorg passed away peacefully, 
at his home in Jena, at the age of 45 years.  
Marcus Ansorg was born on 18th December 1970, in Arnstadt, in the former East Germany. After studying physics at the Friedrich Schiller University of Jena (1990-1994), he obtained a Master of Science in Applied Mathematics at Queen Mary University in London (1995) and was awarded the Lionel Cooper Prize in Mathematics. He completed his Ph.D. work with Gernot Neugebauer in Jena on {\it Timelike geodesic motions in the general-relativistic gravitational field of a rigidly rotating disk of dust} (1998), for which he received the dissertation prize of the Friedrich Schiller University.
Marcus  spent the following years at the Institute of Theoretical Physics (Jena, Germany), the Center for Gravitational Physics and Geometry (Pennsylvania State University, USA), the Max Planck Institute for Gravitational Physics (Albert Einstein Institute, Potsdam, Germany), and at the Helmholtz Center (Munich, Germany). While still in Jena, he developed novel numerical methods for the solution of the Einstein field equations with applications to rotating neutron stars and black holes. Marcus used the essence of the equations to be solved to introduce particularly appropriate coordinate transformations in his spectral methods, thus improving the achievable accuracy by several orders of magnitude over previous methods. Later on, the method was used to fully explore the solution space of differentially rotating neutron stars and strange stars in general relativity and to obtain highly accurate initial data for dynamical evolution of such objects. 
Exploiting the great potential of the newly developed numerical methods and ingenious coordinate mappings, Marcus was able to explore the richness of the space of solutions of rotating and self-gravitating bodies in general relativity. Many of the results obtained in this way, which cover fluid bodies but also black holes, have been collected in an important monograph, {\it Relativistic Figures of Equilibrium} (CUP, 2012), which is a perfect match, in beauty and rigor, to the classical work of Chandrasekhar’s {\it Ellipsoidal Figures of Equilibrium} in Newtonian gravity.
A further highlight during these years was his work on initial data for black holes, resulting in one of the most used data sets of its kind in numerical general relativity (e.g., the TwoPuncture code in the Einstein Toolkit). 
Besides his artful mastery of numerical techniques, he also achieved important results with analytical methods, in particular in the context of universal properties of black hole-matter configurations.
In 2010, Marcus  returned to the Friedrich Schiller University in Jena as Professor of Theoretical Physics / Theory of Gravitation. He was an enthusiastic lecturer and advisor, and his love of science and his productivity in general relativity remained undiminished. During this last period, Marcus shared not only technical knowledge with the members of his group, but also his own principles and philosophy toward the numerical methods employed in the solution of equations. He was not only an advisor, but also a mentor. Not a boss, but a real friend. 
In recent years he also successfully applied his numerical methods in quantum field theory and quantum gravity and his ideas will remain alive in future work. Moreover, together with colleagues from Potsdam and Jena, Marcus developed the novel method of “fully pseudospectral time-evolution” to study time-dependent processes with very high accuracy.
Besides being an outstanding physicist and mathematician, Marcus was a talented athlete and his capabilities appeared unlimited to anyone practicing sports with him. He derived great pleasure from music. He enjoyed singing and playing the guitar. He was also a lover of nature and wilderness, which he would explore either with his (actually Bernd Schmidt's) kayak or on hiking trips with the company of his wife and two sons and friends. He valued family and friendship deeply and was always quick to lend his sincere and well thought through comments, but also his particular humor and his warm smile to any conversation.
Marcus had a sharp mind. For those of us who had the chance to sit down with him at the work table after having struggled at length with a particularly tough problem, there was a feeling of fascination and incredulity when he would produce a neat, simple but subtle answer in a very short time. It invariably looked like magic.
Marcus was systematic. He had always a clear picture in mind of the detailed steps we wanted to take, from {\it a} to {\it z}. This feature, sometimes mistakenly perceived as some kind of rigidity, was rather the methodological key that provided the necessary conditions for deploying a truly audacious research attitude: Marcus loved to finish a meticulous article with a bold conjecture. Then he would try to prove or assess it in a subsequent work. This fresh attitude had impressive pay-offs, as illustrated by the beautiful area-angular momentum-charge inequalities that he proved for black hole horizons surrounded by matter or the elegant equalities involving the product of event and Cauchy horizon areas of stationary black holes, establishing a surprising and largely unattended link to black hole entropy insights in the supergravity context.
Marcus had a sensitive soul. This had so many manifestations, some of them apparent, some of them hidden. 
Ranging from his singular appreciation of his beloved general relativity to the grace of his voice over his guitar. 
And Marcus was a loyal friend. This was part of his characteristic and profound sense of honor. Often this quality did not make his life easier, but it always made it more beautiful and meaningful. And ours too. 

Marcus died after a severe illness, which prematurely ended his remarkable life and career. We mourn with his family the loss of a wonderful person and will honor and cherish his memory.

\vfill\eject

\section*{\centerline
{EGM20}}
\addtocontents{toc}{\protect\medskip}
\addtocontents{toc}{\bf Conference Reports:}
\addcontentsline{toc}{subsubsection}{
\it  EGM20, by Abhay Ashtekar}
\parskip=3pt
\begin{center}
Abhay Ashtekar, Pennsylvania State University 
\htmladdnormallink{ashtekar-at-gravity.psu.edu}
{mailto:ashtekar@gravity.psu.edu}
\end{center}

The 20th Eastern Gravity Meeting was held Friday June 9 and Saturday June 10, 2017 at Penn State University. EGM went very well. We had about 72 participants and there were 47 talks of 15 minutes each.\hfill \\
See 
\htmladdnormallink 
{\protect {\tt {http://gravity.psu.edu/events/egm20/egm20-program.shtml}}}
{http://gravity.psu.edu/events/egm20/egm20-program.shtml}
for a complete listing of speakers and talks.
Since the LIGO collaboration had announced the the third event just a week before EGM20, the meeting began with three talks on this event and we also had two after dinner talks on Gravitational waves. In the first, the audience heard a 5 minute original music piece that was prepared by a Penn State undergraduate (Daniel George) in collaboration with a music professor based on the actual spectrum of the first LIGO event. The second was a short talk on the Future Plans for Gravitational Wave Science. These events were very well received.

Several senior participants commented that they were impressed by the overall high quality of student and post-doc presentations. I fully concur. The judges concurred on two winners of the prize for best student talk: \hfill \\
Beatrice Bonga for the talk \hfill \\
 {\it On the Conceptual Confusion in the Notion of Transverse-Traceless Modes}, \hfill \\ 
and Rahul Kashyap for the talk  \hfill \\
{\it Type Ia Supernovae through Spiral Instability in Binary White Dwarf Mergers}.

\end{document}